\documentclass[prl,floatfix,amsmath,amsfonts,twocolumn]{revtex4}  
\usepackage{amsmath}
\usepackage{graphicx}
\usepackage{graphics}
\usepackage{epsfig}
\usepackage{bm}
\usepackage{color}

\newcommand{\T}{{\cal T}}
\newcommand{\cA}{{\cal A}}
\newcommand{\cU}{{\cal U}}

\newcommand{\cE}{{\cal E}}

\newcommand{\cK}{{\cal K}}

\newcommand{\tk}{\tilde{k}}

\newcommand{\bPsi}{\bm{\psi}}

\newcommand{\bphi}{ \mbox{\boldmath$\phi$\unboldmath}}
\newcommand{\bsigma}{ \mbox{\boldmath$\sigma$\unboldmath}}
\newcommand{\bvarphi}{ \mbox{\boldmath$\varphi$\unboldmath}}

\newcommand{\ba}{{\bm a}}
\newcommand{\bb}{{\bm b}}
\newcommand{\bk}{{\bm k}}
\newcommand{\bu}{{\bm u}}
\newcommand{\br}{{\bm r}}
\newcommand{\bs}{{\bm s}}
\newcommand{\bA}{{\bm A}}

\newcommand{\bP}{{\bm P}}
\newcommand{\rA}{{\rm A}}

\newcommand{\rP}{{\rm P}}
\newcommand{\bxi}{{\bm \xi}} 
 
\newcommand{\rev}[1]{\textcolor{black}{#1}}
 
 \newcommand{\ES}[1]{\textcolor{black}{#1}}
 
\usepackage{ulem}

\begin{document}

\title{Solitons in inhomogeneous gauge potentials: integrable and nonintegrable dynamics}

\author{Y. V. Kartashov$^1$, V. V. Konotop$^2$, M. Modugno$^{3,4}$, and E. Ya. Sherman$^{4,5}$}
\affiliation{
$^1$Institute of Spectroscopy, Russian Academy of Sciences, Troitsk, Moscow, 108840, Russia
	\\
 $^2$Departamento de F\'{i}sica and Centro de F\'{i}sica Te\'orica e Computacional, Faculdade de Ci\^encias, Universidade de Lisboa, Campo Grande, Ed. C8, Lisboa 1749-016, Portugal 
 \\
 $^3$Department of Theoretical Physics and History of Science, University of the
	Basque Country UPV/EHU, 48080 Bilbao, Spain
\\
$^4$IKERBASQUE Basque Foundation for Science, 48013 Bilbao, Spain
\\
$^5$Department of Physical Chemistry, The University of the Basque Country
	UPV/EHU, 48080 Bilbao, Spain}

\date{\today}

\begin{abstract}
We introduce an exactly integrable nonlinear model describing the dynamics of spinor solitons 
in  space-dependent \rev{matrix} gauge potentials of rather general types. 
The model is shown to be gauge equivalent to the integrable system of vector nonlinear 
Schr\"odinger equations known as the Manakov model. As an example we consider a 
self-attractive Bose-Einstein condensate with random spin-orbit coupling (SOC). 
If Zeeman splitting is also included, the system becomes nonintegrable. We illustrate 
this by considering the random walk of a soliton 
in a disordered SOC landscape. While at zero Zeeman splitting the soliton moves without 
scattering along linear trajectories in the random SOC landscape, at nonzero splitting 
it exhibits strong scattering by the SOC inhomogeneities. For a large Zeeman splitting 
the integrability is recovered. In this sense the Zeeman splitting serves as a parameter controlling the crossover between two different integrable limits. 
   
\end{abstract}

\maketitle

Gauge invariance having its origins in the theory of electromagnetism, is known to be a general principle playing crucial role in  almost any field~\cite{history}. One of its applications, intensively discussed nowadays are the synthetic gauge fields 
and potentials. Such potentials of practically arbitrary form can be created in atomic systems, illuminated by proper combination of the laser 
beams~\cite{Ruseckas}. In this way it is possible to emulate in systems of neutral atoms analogs of electric~\cite{electric} and magnetic~\cite{magnetic} fields, 
as well as the spin-orbit coupling (SOC). The latter technique   
has recently made possible to engineer spin-orbit coupled Bose-Einstein condensates (SO-BECs)~\cite{Nature}. Two important properties, 
the tunability  of the SOC in atomic systems~\cite{tun-SO1,tun-SO2,tun-SO3,tun-SO4}, as well as the  intrinsic nonlinearity of SO-BECs stemming from 
inter-atomic interactions, have stimulated extensive studies of soliton dynamics in BECs with inhomogeneous SOC. In particular, 
the interactions of one-dimensional (1D) solitons in SO-BEC with a localized coupling defect have been studied in 1D~\cite{KarKonZez14} and in 2D~\cite{Malomed_2Ddefect} settings, and the propagation of soliton 
in a BEC with inhomogeneous helicoidal SOC was addressed~\cite{KarKon2017}.

The gauge invariance is also known to be a powerful tool of generating and studying nonlinear integrable 
systems~\cite{FadTak}. In particular, two integrable nonlinear equations describing different physical phenomena, may be found to be 
gauge equivalent, i.e. reducible to each other by a gauge transformation. Also, gauge transformation can be used to generate 
continuous~\cite{cont-gauge} and discrete~\cite{KonChubVaz-1,KonChubVaz-2} integrable models with inhomogeneous coefficients departing from the homogeneous ones.    

It was found previously~\cite{KarKon2017} that if a quasi-1D BEC with equal inter-- and intra--component interactions  has a helicoidal structure, and no other potentials or Zeeman splitting is present, the dynamics of soliton  is reduced to the soliton of the exactly integrable Manakov model~\cite{Manakov} which is a system of nonlinearly coupled SU(2) invariant nonlinear Schr\"odinger (NLS) equations. A similar result was pointed out in~\cite{KarKonZez14} for the case of a particular inhomogeneous gauge potential. The possibility of generalizing these results for arbitrary   potentials still remains an open question.  

{ In the present Letter we prove that coupled NLS equations with $x-$dependent \rev{matrix} Hermitian gauge potential of a general type, is an integrable model, which is gauge equivalent to the Manakov system. The} 
inclusion of the Zeeman splitting, quantified below by the field $\Omega$, makes the system non-integrable at 
finite values of $\Omega$, while its integrability is restored in the limit $\Omega\to\infty$. The strength of the  Zeeman field is a parameter describing a crossover between two different integrable limits.  

To study the above  mentioned crossover, we address the evolution of a matter soliton in a BEC with a random SOC (for recent study of random SOC in linear systems see \cite{randSOC}). This is the second goal of this Letter. In particular, we show that the gauge transformation in the integrable case effectively separates random evolution of the pseudo-spin and deterministic evolution of the soliton envelope with random initial 
conditions, similarly to transformation between random and regular SOC explored in the linear theory~\cite{gauge-linear}. The Zeeman field couples these dynamical processes leading to anomalous diffusion of a soliton.  

Let us consider a 1D Gross-Pitaevskii equation (GPE)  describing a spinor $\bPsi(x,t)=(\psi_1,\psi_2)^T$ 
($T$ stands for transpose) in the presence of a random  gauge 
potential $\rA(x)$ and of a Zeeman coupling $\Omega\sigma_{3}/2$:
\begin{equation}
\label{GPE}
i\frac{\partial \bPsi}{\partial t}=\frac{1}{2}\rP^2(x)  \bPsi+\frac{\Omega}{2}\sigma_3\bPsi-\left(\bPsi^\dag\bPsi\right)\bPsi.
\end{equation} 
Here $\rP=-i \partial_x+\rA(x)$ is a generalized momentum, and $\sigma_{1,2,3}$ are the  
Pauli matrices. Inter- and intra-species interactions are assumed to be attractive and equal.
The units are chosen to make the atomic mass $M=\hbar=1$, and $\bPsi$ is normalized to have nonlinear coefficient equal to 1. 
  
We impose three constraints on the $x$-dependent gauge field, requiring it to be Hermitian  $ \rA=\rA^\dagger$, 
 { as needed} for Hermiticity of the generalized momentum, to anti-commute with 
the time reversal operator for spin $1/2$ particles $\T=i\sigma_2\cK$, where $\cK$ is the complex 
conjugation: $\rA\T+\T\rA=0$, and to have  determinant equal to a constant, characterizing the SOC strength: $\det\rA=-\alpha^2$. 
The first two requirements define the general  { form} of the gauge field:
$\rA(x)= \bsigma \ba$,  where $\ba=(a_1,a_2,a_3)$ is a real vector and $\bsigma=(\sigma_1,\sigma_2,\sigma_3)$ is the Pauli matrix vector. 
  Next we consider an eigenstate ${\bm \xi}_k(x)$ of the $\rP$ operator:  $\rP{\bm \xi}_k=k{\bm \xi}_k$, $k$ being the eigenvalue. We also define $\bphi_1(x)=e^{-ikx}{\bm \xi}_k$  and  $\bphi_2=\T\bphi_1$.   It is straightforward to verify that 
 $  \rP\bphi_{j}=0$ and $\bphi_{j}^\dagger\bphi_j$ is $x$-independent. Thus the spinors $\bphi_{1,2}(x)$ make up an orthonormal basis in the spinor subspace: $\bphi_{i}^\dagger\bphi_j=\delta_{ij}$. The vectors $\bphi_1$ and $\bphi_2$ describe opposite pseudo-spin distributions $\tilde{\bsigma}(x)$ and $-\tilde{\bsigma}(x)$, where $\tilde{\bsigma}=(\tilde{\sigma}_1,\tilde{\sigma}_2,\tilde{\sigma}_3)$ with $\tilde{\sigma}_j(x)=\frac{1}{2}\bphi_1^\dagger\sigma_j\bphi_1$ $(j=1,2,3)$.
 
Using the basis $\bphi_{1,2}(x)$ one can write the solution of (\ref{GPE}) as
$
\bPsi(x,t)=u_1(x,t)\bphi_1(x)+u_2(x,t)\bphi_2(x)
$, 
 and verify that the ``envelope" spinor  $\bu=(u_1,u_2)^T$ solves the equation
  \begin{equation}
  \label{Manakov_gen}
  i\bu_t+\frac{1}{2}\bu_{xx}+(\bu^\dagger\bu)\bu=\Omega \kappa(x)\bu ,\quad  \kappa=\left(\begin{array}{cc}
  s_3 & \kappa_3
  \\
  \kappa_3^* &-s_3
  \end{array}\right),
  \end{equation}
  where $\kappa_3(x)=\frac{1}{2}\bphi_1^\dagger\sigma_3\bphi_2$ describes the coupling of the envelope components.  This allows us to interpret the vectors $\bphi_{1,2}$ and $\bu$ respectively as the pseudo-spin and 
  the soliton ``degrees" of freedom, which are coupled by the Zeeman field when $\Omega\neq0$. In the absence of Zeeman field, $\Omega=0$, Eq. (\ref{Manakov_gen}) describes deterministic evolution of the envelope $\bu$.  We note, that even in the case of deterministic initial condition for 
  the field $\bPsi_0(x)=\bPsi(x,0)$, the initial conditions for the fields $u_{1,2}(x,t)$ are random: they are defined by the projections of $\bPsi_0(x)$ on $\bphi_{1,2}(x)$. 
  
  \rev{Remarkably, the described separation of spinor and nonlinear degrees of freedom can be performed also in 2D and 3D cases for spatially dependent non-Abelian gauge potentials, whose components are related by zero-curvature conditions~\cite{suppl}.}
  
 For  $\Omega=0$ Eq.~(\ref{GPE}) is
  gauge equivalent to the Manakov model ~\cite{Manakov}, hence it is \textit{exactly integrable}.  
  Indeed, for $\Omega=0$, Eq.~(\ref{GPE}) is obtained from the compatibility condition $U_t-V_x+[U,V]=0$ of the eigenvalue problems:  
  \begin{eqnarray}
  \label{eigen_prob}
   \bvarphi_x=U\bvarphi-i\lambda \bvarphi\cE, \quad \mbox{with} \,\, U=i(\lambda \cE+\cA+\cU)
   \end{eqnarray}
  and  $\bvarphi_z=V\bvarphi-i\lambda^2 \bvarphi\cE$, 
  with $V=i\lambda^2\cE+i\lambda\cU+\frac{1}{2}\cE \cU_x-\frac{i}{2}\cE\cU^2-\frac{i}{2}\cE[\cA,\cU] $, 
  where  $\bvarphi$ is a $3\time 3$ matrix, $\lambda$ is the spectral parameter, $\cE=$diag$(1,1,-1)$, 
  \begin{equation} 
  \cA= \left( 
  \begin{array}{ccc}
  a_3 & a_1+ia_2 & 0
  \\
  a_1-ia_2 & -a_3  & 0
  \\
  0 & 0 & 0
  \end{array}
   \right)\!,\,\,
  \cU= \left( \begin{array}{ccc}
  0 & 0 & \psi_1^* 
  \\
  0 & 0  & \psi_2^* 
  \\
  \psi_1 &\psi_2 & 0
  \end{array}\right).
  \end{equation}
  When all $a_j=0$ we recover the $UV-$representation of the Manakov model~\cite{Doctor}.
  
  Turning now to the opposite limit of large Zeeman splitting  and performing the rotation ${\bm \Psi}=e^{-i\Omega\sigma_3t/2}{\bm \psi}$~\cite{KarKonZez14}, 
  the GPE (\ref{GPE}) for the spinor ${\bm \Psi}$ preserves its original form, but now without Zeeman field and with time-dependent 
  gauge potential  $\rA_\Omega(t)=a_3\sigma_3+ e^{i\Omega\sigma_3t}(a_1\sigma_1+a_2\sigma_2)$. At $\Omega\to\infty$, the last 
  components of $\rA_\Omega(t)$ become rapidly oscillating and their average effect on the dynamics vanishes. This corresponds 
  to the rotating wave approximation with SOC being a perturbation with respect to the Zeeman field. 
  In this case $\rA_\Omega(t)\to a_3\sigma_3$ and thus the model again becomes exactly 
  integrable (although different from the limit of zero Zeeman splitting). This limit can be also viewed as the nonlinear 
  analog of Paschen-Back effect~\cite{Paschen-Back}, which for an atom with a random coupling 
  (although not having a gauge structure) was discussed in~\cite{MoSherKon17}.  

At $\Omega > 0$, the system is not integrable, but a wavepacket obeys the Ehrenfest theorem~\cite{Ehrenfest}  
 \begin{equation}
  \label{Newton}
  \frac{dX}{dt}=\Pi, \quad \frac{d\Pi}{dt}= i\frac{\Omega}{2\|\bPsi\|^{2}}\int\bPsi^\dagger[\sigma_3,\rA]\bPsi dx\,(\equiv F(t))
  \end{equation}
  which is written in terms of  the soliton center of mass position
   $
  X=\|\bPsi\|^{-2}\int_{-\infty}^{\infty}\bPsi^\dagger x\bPsi dx
   $,
 where the norm  
  $
  \|\bPsi\|^{2}=\int_{-\infty}^{\infty}\bPsi^\dagger \bPsi\, dx
    $
  is a conserved quantity, and of
  the integral momentum of the soliton
    $
  \Pi=\|\bPsi\|^{-2}\int_{-\infty}^{\infty}\bPsi^\dagger \rP(x) \bPsi\, dx
  $
   which is a conserved quantity in both the integrable limits discussed above. 

To explore the crossover between the integrable limits  we consider a soliton in a BEC with 
 SOC of the form $\rA(x)=\alpha \sigma_1e^{i\sigma_3\theta(x)}$, where $\theta(x)$ is a random function. The experimental feasibility of the model stems from different scales of the wavelength of the laser beams producing the SOC (typically below one micron), and of the random potential variations, which is about 10 $\mu$m for a 1D condensate of a transverse width of a few microns. The random field can be produced by spatially modulated beams, as shown in~\cite{suppl} for an example of a tripod scheme~\cite{Ruseckas}. Use of monochromatic quasi-nondiffracting beams~\cite{nondiff1} allows for designing practically arbitrary spatial modulations~\cite{nondiff5} on basis of algorithms developed in~\cite{nondiff3}. \rev{For alternative possibilities of producing prescribed gauge potentials see e.g.~\cite{GaugeReview}.}

For the numerical simulations we choose the random 
function $\theta(x)=2\pi f(x)/f_\textrm{max}\in[-2\pi,2\pi]$, where $f(x)=\sum_{j=-n}^nr_je^{-(x-j-r_j)^2/2}$, with $r_j\in [-0.5,0.5]$ 
being a uniformly random distribution with zero average value,  $\langle \theta(x)\rangle=0$ (angular brackets stand for statistical averaging). 
The initial condition in all simulations was chosen in the form of a wavepacket with only lower state populated 
$\bPsi_0(x)=e^{i v x}\mbox{sech}(x) (0, 1)^T$, which corresponds to $\|\bPsi\|^{2}=2$, $X(0)=0$, $\Pi(0)= v $, and $F(0)=0$. The evolution of such state was 
obtained by solving Eq. (\ref{GPE}) 
for long times (up to $t_{f}=10^3$) 
 for each realization of $\theta(x)$, and the subsequent averaging was performed over $10^3$ realizations of $\theta(x)$.

\begin{figure}[ht]
	\centering
	\includegraphics[width=\columnwidth]{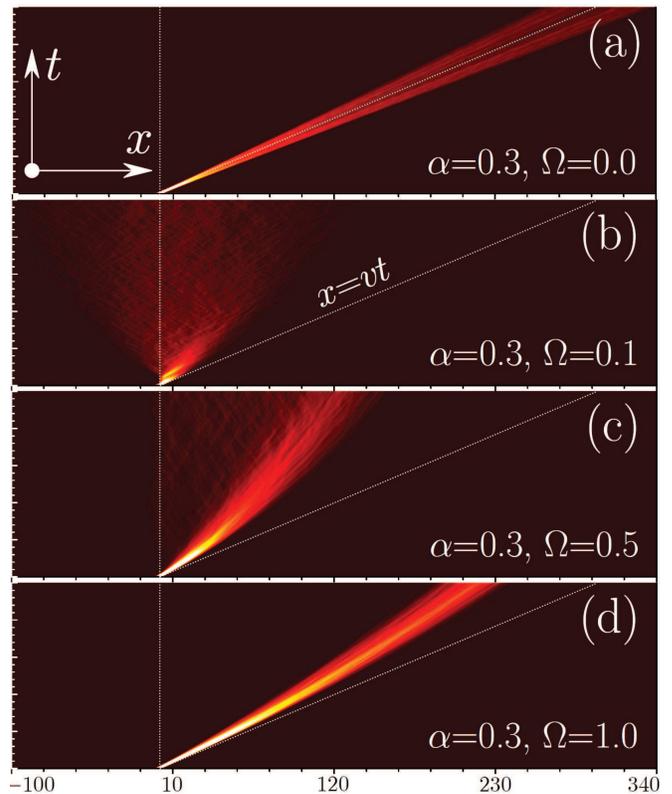}
	\caption{Evolution of the average density of the dominant 
	component $\langle |\psi_2|^2\rangle$ for $\alpha=0.3$, $ v =0.3$  
	in the initial condition $\bPsi_0$, and different strengths of Zeeman field.  Distributions are shown up to $t =10^3$. 
	 Dotted lines indicate the initial wavepacket position (vertical line) and center trajectory in the integrable case $\Omega=0$ (oblique line). 
	}
\label{fig:one}
\end{figure}

 The evolution of the averaged atomic density of the dominant $\psi_2$ component is illustrated in Fig.~\ref{fig:one}. 
For each realization of the random function $\theta(x)$ the excited soliton moves as a localized object that does not spread, i.e. the $\psi_1$ component always accompanies the dominant $\psi_2$ component, 
and moves along the same trajectory in the $(x,t)$ plane. Such individual trajectories are resolvable in the averaged 
density distributions. The dynamics in panels (a) to (d) shows the crossover between the two integrable limits 
of $\Omega=0$ and $\Omega\to\infty$ (the transition to the latter limit is obvious already at $\Omega\sim 1$). A peculiarity of this system is that even in the integrable limit $\Omega=0$ 
we observe a small divergence of the linear trajectories from the central one, indicated by the dashed line 
[Fig.~\ref{fig:one}(a)]. This reflects the fact that the eigenvalue problem (\ref{eigen_prob}) is random even for 
deterministic initial conditions, i.e. solitons generated by the same initial condition ${\bm\psi}_{0}$ in different 
realizations of the gauge potential acquire randomly distributed parameters, including random velocities 
(concentrated in a narrow interval around $ v $). 
Already for small $\Omega\sim 0.1$, when the integrability is lost [panels (b) and (c)], one observes a considerable 
scattering of solitons by inhomogeneities of the SOC landscape that in many cases 
may lead to the inversion of the soliton velocity. This scattering occurs due to random perturbation 
in the right-hand side of (\ref{Manakov_gen}). In terms of the ``Newtonian" 
dynamics of the soliton (\ref{Newton}), this is the effect of the time dependent force $F(t)$ stemming from the noncommutativity 
of the gauge and Zeeman fields. Scattering becomes much weaker at $\Omega\sim 1$ [Fig.~\ref{fig:one}(d)].

To characterize the statistical properties of the evolution dynamics, we studied the average soliton 
displacement and the mean squared displacement (MSD). 
In all realizations of SOC landscape the integral soliton center of mass $X(t)$ practically 
coincides with the position of the soliton maximum $x_{\rm m}(t)$ defined through the relation  
$S_0(t)\equiv\bPsi^{\dagger}(x_{\rm m},t)\bPsi(x_{\rm m},t)=\max_x[\bPsi^{\dagger}(x,t) \bPsi(x,t)]$. However, definition for 
the MSD based on the position of maximum $d= \langle x_{\rm m}^2\rangle -\langle x_{\rm m}\rangle^2$ is much more 
accurate than the integral one, because it disregards radiation emitted by soliton interacting with random potential. For the above reasons, below we use averaged quantities based on the position of soliton maximum $x_{\rm m}(t)$.  

\begin{figure}[h]
	\centering
	\includegraphics[width=\columnwidth]{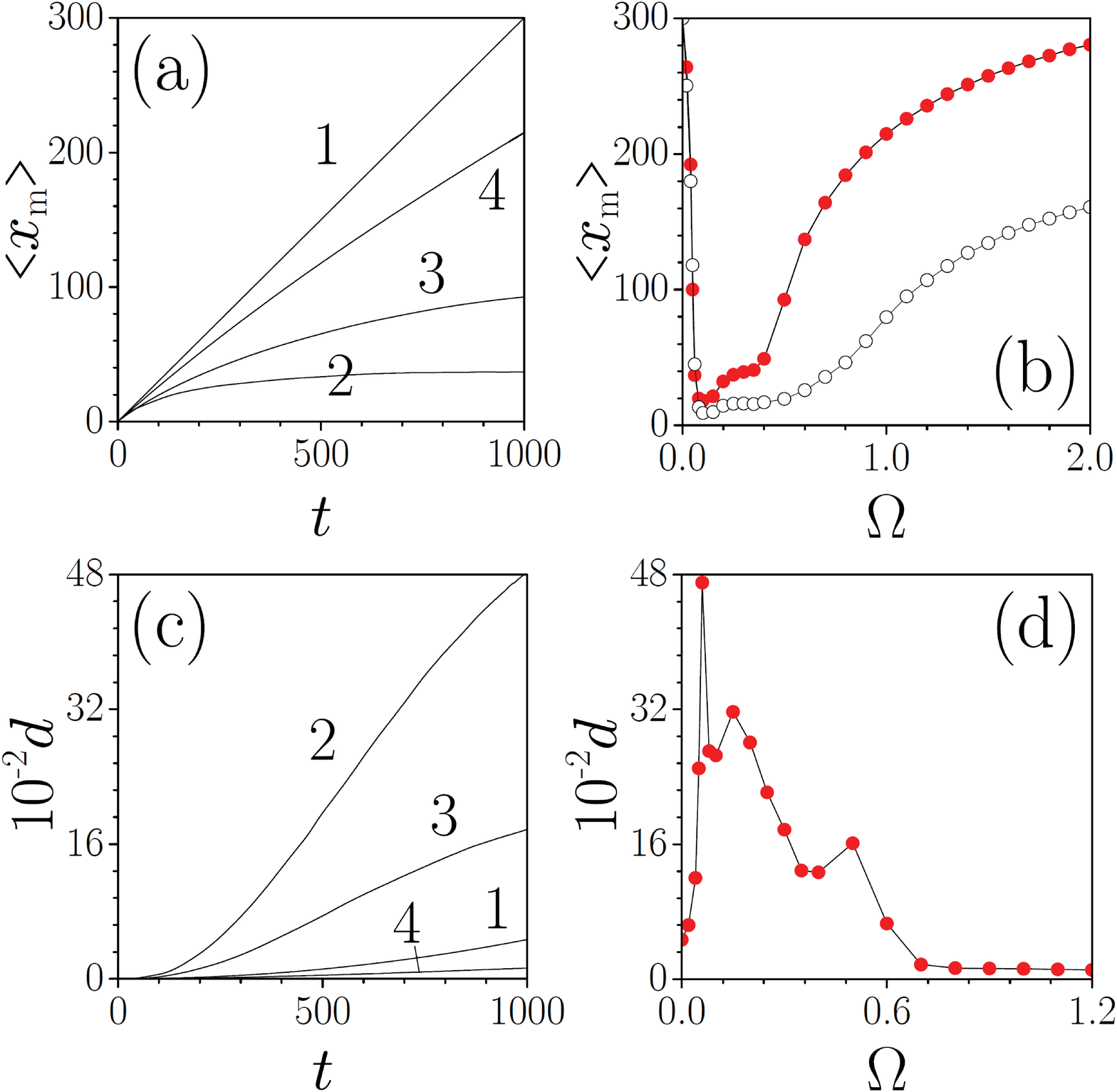}
	\caption{(a) Dynamics of the averaged soliton center for $\Omega=0$ (curve 1),
		0.06 (curve 2), 0.5 (curve 3), and 1 (curve 4). (b) Averaged displacement at $t=10^3$ vs Zeeman splitting $\Omega$ {(red dots $\alpha=0.3$, white dots $\alpha=0.42$)}. 
		(c) The mean squared displacement for $\omega=0$ (curve 1), 0.06 (curve 2), 0.3 (curve 3), and 1 (the lowest curve 4). 
		(d) The mean squared displacement at $t =10^3$ vs Zeeman splitting $\Omega$. In all cases $ v =0.3$, $\alpha=0.3$. 
	}
	\label{fig:two}
\end{figure}

The averaged displacement and MSD are shown in Fig.~\ref{fig:two}. Panel (a) shows variation with time of the averaged 
displacement in the crossover between the two integrable limits (the displacement first rapidly decreases at $\Omega\sim 0.1$, curve 2, 
but then gradually increases with grows of $\Omega$, see curves 3 and 4). The effect of the Zeeman field $\Omega$ is illustrated in 
Fig.~\ref{fig:two}(b), where a deep minimum appears in the displacement computed at $t_{f}=10^3$ {obtained for two different Zeeman fields}. 
This minimum, observed when the Zeeman field and the strength of the gauge field are of the same order, $\Omega\sim \alpha$, 
corresponds to a parameter range where the impact of the effective force $F(t)$ on the soliton propagation is strongest. 
In Fig.~\ref{fig:two}(c) we show the anomalous diffusion of the soliton (recall that the parameter $d$ characterizes 
the deviation of trajectories of the soliton motion from mean trajectory, i.e. in a sense this is a 
measure of the width of the averaged patterns from Fig.~\ref{fig:one}). In the integrable limit $\Omega=0$, the 
curve 1 represents an exact parabola, because now both $\langle x_{\rm m}^2\rangle$ and $\langle x_{\rm m}\rangle^2$ scale as $t^2$, 
with the coefficients of the proportionality being determined by the distribution of the discrete spectrum of the 
eigenvalue problem (\ref{eigen_prob}). Much stronger diffusion is observed in the nonintegrable limit at weak 
Zeeman field (curves 2 and 3). Interestingly, the anomalous diffusion becomes weaker with the increase of $\Omega$, 
and it may be even lower than diffusion at $\Omega=0$. This is also obvious from Fig.~\ref{fig:one}(d), where the 
width of the pattern becomes relatively narrow. This is the effect of the fast rotations, leading to zero effective
gauge potential $\rA_\Omega (t)\to 0$ (see above) for the chosen model of SOC. Thus in our system MSD is also 
nonmonotonic function of the Zeeman field: it is very small in two integrable limits and acquires maximal values in 
the crossover regime [Fig.~\ref{fig:two}(d)].

\begin{figure}[h]
	\centering
	\includegraphics[width=\columnwidth]{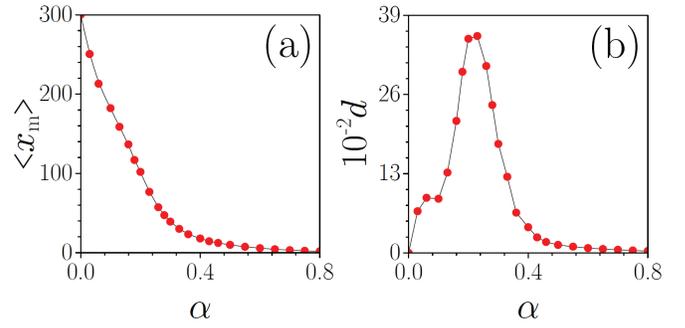}
	\caption{Averaged displacement of the soliton center (a), and the MSD vs SOC strength $\alpha$ (b), at $t=10^3$, for $ v =0.3$ and $\Omega=0.3$.
		 }
	\label{fig:three}
\end{figure}

It follows from (\ref{Newton}) that for sufficiently small $\Omega$ one can estimate $F(t)\sim\alpha\Omega$, i.e. by fixing a nonzero Zeeman field and increasing the SOC strength one results in stronger effect of the random gauge potential on the soliton. The decay of the force at large $\Omega$ is due to fast oscillations in the integrand in (\ref{Newton}), corresponding to the limit of rotating wave approximation.  Quantitatively this  is illustrated in Fig.~\ref{fig:three}, where the average soliton displacement rapidly decreases to zero 
(due to increasing dispersion of the soliton trajectories)
and by a sharp maximum of the MSD in the region where $\alpha\sim\Omega$ [cf. Fig.~\ref{fig:two} (d)]

\begin{figure}[ht]
	\centering
	\includegraphics[width=\columnwidth]{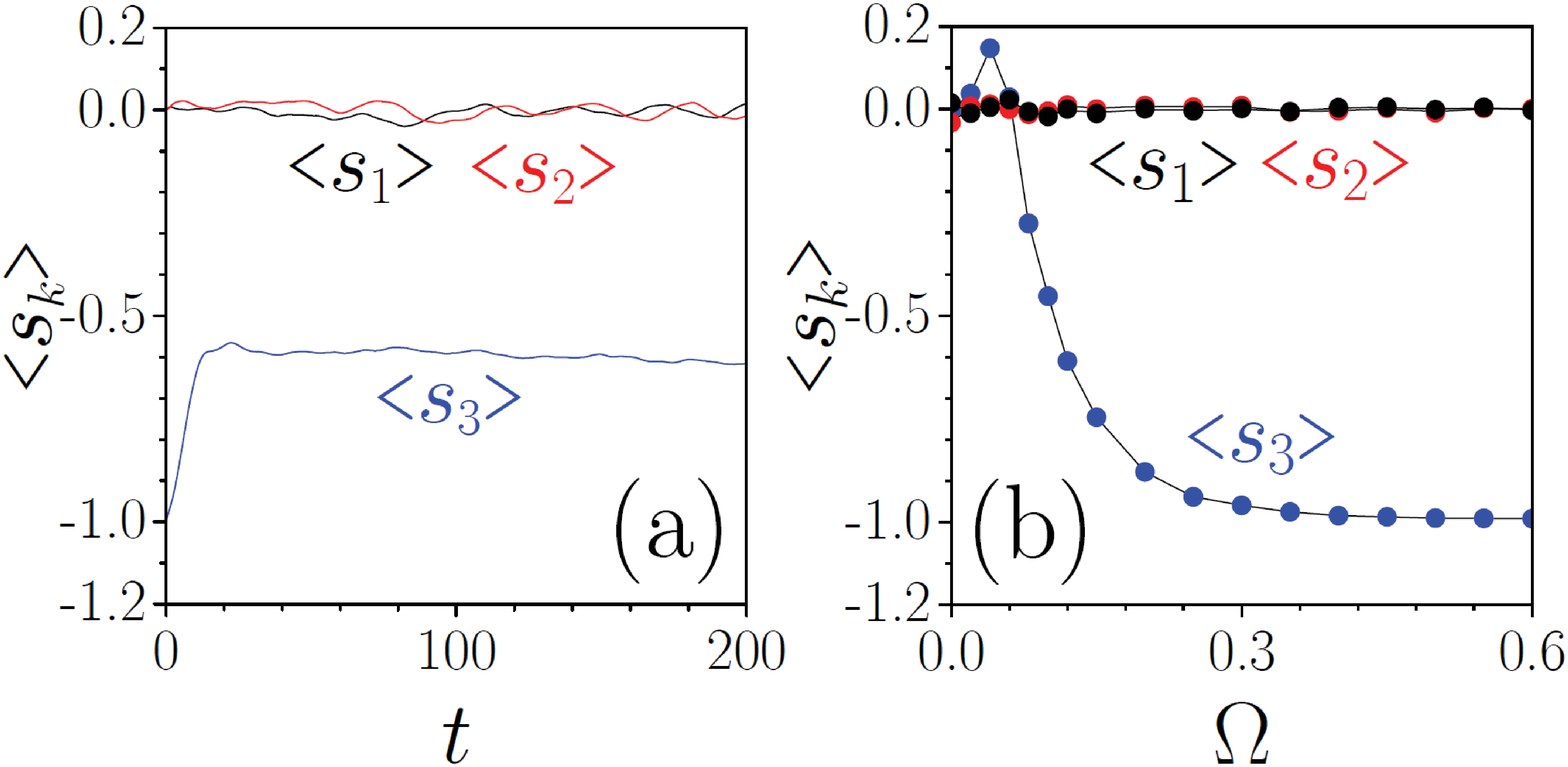}
	\caption{(a) Evolution of averaged pseudo-spin components at $\Omega=0.15$. 
	(b) Final averaged pseudo-spin components at $t =10^3$ vs Zeeman splitting. In all cases  $ v =0.3$ and $\Omega=0.3$.
	}
	\label{fig:four}
\end{figure}

Turning to the (pseudo-)spinor characteristics we define 
$
\bs(t)=
{S_{0}^{-1}(t)}
\bPsi^{\dagger}(x_{\rm m},t)\bsigma \bPsi(x_{\rm m},t)
$
 with
$\bs(t)$ being always on the Bloch sphere: $|\bs|=1$. 
The choice of the initial $\bPsi_0(x)$ for numerical simulations corresponds 
to the ``pure'' state soliton bearing the spin: $\bs(0)=(0,0,-1)$. Due to random time-dependence of 
the direction of $\bs$, determined by the realization of the random gauge field, the ensemble-averaged
$\langle s_{3}\rangle$ undergoes a 
relatively fast relaxation [see the example in 
Fig.~\ref{fig:four} (a)], characterized by time \cite{Glazov} of $\tau_{\rm s}\sim 1/\alpha^{2}\sqrt{\langle\Pi^{2}\rangle}\zeta,$
[cf. Eq.~(\ref{Newton})] with $\zeta$ being the correlation length of the $\rA(x)-$field. 
For the chosen model parameters $\zeta\sim 1$  and  $v=\alpha=0.3,$
we obtain $\tau_{\rm s}\sim 30,$ in a good agreement with Fig.~\ref{fig:four}(a). 
The maximal relaxation of the initial spin is achieved in the integrable limit at zero 
Zeeman splitting [Fig.~\ref{fig:four}(b)].  A specific feature of the nonintegrable regime, shown in Fig. \ref{fig:one},
is the decrease in $\sqrt{\langle\Pi^{2}\rangle}$ with time, slowing the relaxation down, as can be seen in  
Fig.~\ref{fig:four}(a).
Such a behavior is a consequence of the ``independent'' 
deterministic dynamics of the soliton center of mass, described by $\bu$ in (\ref{Manakov_gen}) at
$\Omega=0$ and stochastic dynamics of the soliton pseudo-spin $\bs.$ 
Increasing the Zeeman field results in restoring the pure  character of the soliton spin state, which is 
observed in Fig.~\ref{fig:four} already at $\Omega\gtrsim 0.3$. After a short interval of growth of $\langle s_3 \rangle$, in the interval $0.04\lesssim\Omega\lesssim 0.3$, 
increasing of the Zeeman splitting results in gradual decrease of $\langle s_3\rangle$. 
 	
To conclude, we described the evolution of solitons in inhomogeneous gauge potentials. 
In the absence of the Zeeman field the model is exactly integrable for 
arbitrary spatial distributions of the \rev{matrix} gauge potential. Solitons, and more sophisticated solutions can also be constructed using the Inverse Scattering Technique.  We described statistics of solitons 
affected by random SOC. In the integrable case solitons move with constant velocities which are different for different realization of the SOC. When the Zeeman splitting is large, the system again 
approaches an integrable limit, although different from the one at zero Zeeman splitting. The crossover between these two integrable limits is characterized by strong interaction of a soliton with the random gauge potential, manifesting itself in a slowing down average motion and strongly anomalous diffusion of solitons. Each soliton carries a pseudo-spin.  The  dynamics of ensemble-averaged pseudo-spinors is characterized by two temporal scales: the fast relaxation at initial stages, well-described in quasi-linear approximation, and the long-time slow evolution.

\begin{acknowledgments}
V.V.K. acknowledges support of the FCT (Portugal)
 grants UID/FIS/00618/2013.
M.M. and E.S. acknowledge support by the Spanish Ministry of Economy, Industry, and Competitiveness (MINECO) and the 
European Regional Development Fund FEDER through Grant No. FIS2015-67161-P (MINECO/FEDER, UE), 
and the Basque Government through Grant No. IT986-16.
\end{acknowledgments}


\newpage 

\begin{center}
{\bf Supplemental material}	 
\end{center}
 
\section{On separation of nonlinear time evolution and linear field distribution in a two-dimensional case with a non-Abelian gauge potential.}

The separation of "linear" and "nonlinear" dynamics, reported in the main text for one-dimensional (1D) Gross-Pitaevskii equations (GPE)  can also be performed in the 2D  and 3D cases for specific types of non-Abelian potentials. In order to illustrate this, here we consider 2D GPE
\begin{equation}
\label{GPE_1}
i\frac{\partial \bPsi}{\partial t}=
\frac{1}{2}\bP^2(\br)  \bPsi-\left(\bPsi^\dag\bPsi\right)\bPsi.
\end{equation} 
where 
\begin{eqnarray}
\label{P3D}
\bP=-i\nabla+\bA(\br)
\end{eqnarray}
$\bA=(A_x,A_y)$, with $A_{x,y}$ being $2\times 2$ matrices, is an inhomogeneous non-Abelian gauge potential (it is arbitrary, so far), and the Zeeman splitting is set to zero.

Next we consider the eigenvalue problem
\begin{eqnarray}
\label{eigen1}
\bP\xi_\bk(\br)=\bk\xi_\bk(\br)
\end{eqnarray}
which by the ansatz $\bxi_\bk(\br)=e^{i\bk\cdot\br}\bphi_1(\br)$, is reduced to $\bP\bphi_1=0$, or explicitly 
\begin{eqnarray}
\label{eigen2}
-i\nabla\bphi_1+\bA(\br)\bphi_1=0.
\end{eqnarray}
The obtained equation, is not solvable for arbitrary potential $\bA(\br)$. Indeed considering the components of (\ref{eigen2}) separately one obtains that the condition
\begin{eqnarray}
\label{solvability}
i\frac{\partial A_x}{\partial y}-i\frac{\partial A_y}{\partial x}+[A_x,A_y]=0,
\end{eqnarray}
which can be viewed as the zero-curvature condition, must be satisfied. Notice, that for a constant potential, i.e., for $A_x=$const and $A_y=$const, the solvability condition requires the potential to be Abelian, i.e., to have $[A_x,A_y]=0$.

Thus, we require (\ref{solvability}) to be satisfied. Furthermore, like in the main text, we require the gauge potential to be Hermitian, i.e., $A_{x,y}^\dag=A_{x,y}$ and to anti-commute with the time reversal: $\T\bA+\bA\T=0$. Thus the vector function defined by $\bphi_2=\T\bphi_1$, solves $\bP\bphi_2=0$. Furthermore, it is straightforward to verify that
\begin{eqnarray}
\nabla (\bphi_1^\dagger\bphi_1)=\nabla (\bphi_2^\dagger\bphi_2)=0
\end{eqnarray}
and 
\begin{eqnarray}
\label{orthonorm}
\bphi_1^\dagger\bphi_2=0.
\end{eqnarray}
Now, one can search for a solution of (\ref{GPE}) in a form of the ansatz
\begin{eqnarray}
\bPsi(\br,t)=u_1(\br,t)\bphi_1(\br)+u_2(\br,t)\bphi_2(\br)
\end{eqnarray}
where $u_{1,2}(\br,t)$ are two unknown functions, which solve the equation
\begin{eqnarray}
\label{Manakov_gen_1}
i\bu_t+\frac{1}{2}\nabla^2\bu+(\bu^\dagger\bu)\bu=0
\end{eqnarray} 
where $\bu=(u_1,u_2)^T$ and we used property (\ref{orthonorm}).

The above consideration can be straightforwardly generalized to the 3D case, where the solvability condition for (\ref{eigen2}) requires vanishing of all nondiagonal elements of the curvature tensor.

For the sake of illustration of a (nontrivial) non-Abelian gauge potential for which the separation of nonlinear time evolution and linear spatial spinor field distribution is possible, we consider
\begin{eqnarray}
\label{exampleA}
A_x= \ba(x)\cdot{\bm \sigma}, \qquad A_y=\bb(x)\cdot{\bm \sigma}
\end{eqnarray}
where $\ba$ and $\bb$ depend only on $x$. Then (\ref{solvability}) is reduced to the system of ODEs
\begin{eqnarray}
\label{odes}
\frac{d\bb}{dx}=2\ba\times\bb.
\end{eqnarray}

As it is mentioned above, it follows from (\ref{solvability}) that two arbitrary stationary (coordinate independent) potentials must be Abelian, for the separation ansatz to be applicable. In particular, the conventional two-dimensional Rashba and Dresselhaus couplings are given by coordinate-independent $\ba=(0,1,0),\,\bb=(-1,0,0)$ and 
$\ba=(1,0,0),\,\bb=(0,-1,0),$ correspondingly, with $|\ba\times\bb|=1$. Therefore, these interactions cannot be gauged out \cite{Aleiner}.  However, if either $\ba=\mathbf{0}$, or $\bb=\mathbf{0}$, or the SOC contains the Rashba and Dresselhaus contributions of equal strengths,
then $\ba\times\bb=\mathbf{0},$ and the required gauge transformation is possible~\cite{Tokatly}.

If Eq.~(\ref{odes}) is satisfied by given (either coordinate-dependent or independent) vectors $\ba$ and $\bb$, then for 
the vector potential given by (\ref{exampleA}) we can look for a solution of (\ref{eigen2}) in the form  
\begin{eqnarray}
\bphi_1(\br)={\tilde \bphi}_1(x)e^{-iqy}
\end{eqnarray}
where
\begin{eqnarray}
\frac{1}{i}\frac{d{\tilde \bphi}_1}{d x}+A_x(x){\tilde \bphi}_1=0,
\quad
A_y(x) {\tilde \bphi}_1=q{\tilde \bphi}_1,
\end{eqnarray}
(these equations are consistent) and $q$ is a constant. Finally, looking for solutions of (\ref{Manakov_gen}) independent on $y$ we arrive at the 1D Manakov model considered in the main text.

\section{A scheme for random gauge potentials}
In order to describe a possibility of how a random gauge potential 
can be generated, let us consider a BEC with four-level atoms in a tripod configuration described by the Hamiltonian~\cite{Ruseckas} 
\begin{eqnarray}
H_0=-\hbar(\Omega_1|0\rangle\langle 1|+\Omega_2|0\rangle\langle 2|+\Omega_3|0\rangle\langle 3|) +{\rm h.c.}
\end{eqnarray}
where  $|j\rangle$ ($j=1,2,3)$ are the {low-energy} states and $|0\rangle$ is the excited 
state which is coupled to the states $|j\rangle$ by the Rabi frequencies $\Omega_j$. 
Consider now a BEC \ES{loaded} in a cigar-shaped trap, which is long enough along the $x$-direction 
(say, \ES{approximately}  200$\,\mu$m length)  and has transverse radial width in the $(y,z)-$plane of the
order of $a_\bot\approx 10\,\mu$m.  The coupling of the {low-energy} states with the excited state 
is assured by the two counter-propagating laser beams: 
\begin{eqnarray}
\label{beams}
\begin{array}{c}
\displaystyle{\Omega_{1}=\Omega  e^{i\Theta(\br)} e^{ik_xx+ik_yy}}
\\[3mm] 
\displaystyle{\Omega_{2}=\Omega  e^{i\Theta(\br)} e^{-ik_xx+ik_yy} }
\\[3mm] 
\displaystyle{\Omega_3=2\Omega  e^{ikz} }
\end{array}
\end{eqnarray}
where ${\bm r}=(x,y,z)$, $\vartheta$ is a real constant and $\Omega$ is the field amplitude.  
The beams $\Omega_{1,2}$ \ES{propagating} along the directions $(\pm \cos\varphi, \sin\varphi,0)$, 
where $\cos\varphi=k_x/\tk$  and $\sin\varphi=k_y/\tk$ (i.e. $\tk^2=k_x^2+k_y^2$), 
can be created as superposition of nondiffracting beams \cite{nondiff1}. For example, one can represent
\begin{widetext}
	\begin{eqnarray}
	\label{aux1}
	e^{i\tk \xi}e^{i\Theta(\xi,\eta,\zeta)}= \int_{k_0-\delta k}^{k_0+\delta k}d k_\bot e^{-i\sqrt{k^2-k_\bot^2}\xi} 
	\int_{-\pi}^{\pi}d\nu  {\cal A}(\nu,k_\bot)  e^{ik_\bot (\eta \cos\nu+\zeta\sin\nu )}, 
	\end{eqnarray}
\end{widetext}
where $(\xi,\eta,\zeta)$ are the Euclidian coordinates in the rotated frame, 
$\xi=x\cos\varphi+y\sin\varphi$, $\eta$ and $\zeta$ are the coordinates 
in the plane orthogonal to $\xi-$axis, and $k=\omega/c$. In Eq. (\ref{aux1})  the angular spectrum ${\cal A}(\nu,k_\bot)$ is defined in the Fourier domain,  on  
a narrow annular ring of the width $2\delta k$ having central radius $k_0=\sqrt{k^2-\tilde{k}^2}$ ($\nu$ is the angular coordinate). 

Engineering of the spectrum ${\cal A}(\nu,k_\bot)$ using iterative Fourier methods, 
reminiscent of the methods employed in phase retrieval and image processing 
algorithms \cite{nondiff3}, allows \ES{researchers} to produce quasi-nondiffracting monochromatic 
light patterns with any desired phase or intensity distribution in the $(\eta,\zeta)$ plane and characteristic features with scales $\sim 2\pi/k_0$ ranging from several to hundreds of microns, as demonstrated in \cite{nondiff5}. 

Now, the two dark states of $H_0$ can be found in the form: 
\begin{widetext}
	\begin{eqnarray}
	\label{dark1}
	|D_1\rangle=\frac{1}{\sqrt{2}}\left\{ e^{-ik_xx-ik_yy}|1\rangle-e^{ik_xx-ik_yy}|2\rangle \right\},
	\\[2mm]
	\label{dark2}
	|D_2\rangle=  	\frac{1}{\sqrt{3}} \left\{  e^{-i\Theta(\br)}e^{-ik_xx-ik_yy}|1\rangle+  e^{-i\Theta(\br)}e^{ik_xx-ik_yy}|2\rangle 
	-  e^{-ikz}|3\rangle\right\}.
	\end{eqnarray}
\end{widetext}

The spinor wave-function is {sought} in the form 
\begin{eqnarray}
|\Psi\rangle=\Psi_1 (\br) |D_1\rangle+\Psi_2 (\br) |D_2\rangle. 
\end{eqnarray}
Now, in the absence of interactions the evolution of the spinor $\bPsi=\left(\Psi_1,\Psi_2\right)^T$  
is governed by the Hamiltonian~\cite{Ruseckas}:
\begin{eqnarray}
H_{\rm lin}=\frac{1}{2M}\left(\frac{\hbar}{i}\nabla- A\right)^2+V_{\rm tot}(\br),
\end{eqnarray}
where $M$ is the atomic mass and the 
vector matrix $A$ (known also as Berry connection)  has elements 
\begin{eqnarray}
A_{mn}=i\hbar\langle D_{m}(\br)|\nabla D_{n}(\br)\rangle.
\end{eqnarray}
The total potential $V_{\rm tot}(\br)$ consists of two parts: one is the external trap potential $V_{\rm ext}$ which is a matrix if the components are coupled, while another part $U$ is a matrix potential induced by the laser beams (\ref{beams}). It has components~\cite{Juzel}
\begin{eqnarray}
U_{ij}=\frac{\hbar^2}{2M}\left(\langle \nabla D_i|\nabla D_j\rangle + \sum_{l=1}^{2}\langle D_i|\nabla D_l\rangle\langle D_l|\nabla D_j\rangle\right).
\end{eqnarray}

Due to quasi-one-dimensionality of the condensate, we are interested 
only in the distribution of $\theta(x)=\Theta(\hat{{\bf i}} x)$, i.e. 
in the distribution of $\Theta(\br)$ along the $x$-axis (at $y=z=0$). 
The only requirement for the function $\Theta(\br)$, used so far in (\ref{aux1}), is that 
it must be slowly varying on the scale of the wavelength of the beams $\Omega_{1,2}$, i.e. on the scale  $\lambda=2\pi/k$.

Substitution of the dark states (\ref{dark1}) and (\ref{dark2}) in this formula yields the $x-$component of the dimensionless gauge potential  
\begin{eqnarray}
A= \alpha \sigma_1e^{i\sigma_3\theta(x)}, \qquad \alpha=\sqrt{\frac{2}{3}}a_\bot k_x,
\end{eqnarray}
i.e. the formula used in the text. Here we neglected the derivative of slowly varying $\theta(x)$. 

Considering  the matrix $U$ in the same approximation of slowly varying $\theta(x)$, one obtains that this potential is diagonal: 
\begin{equation}
U=\frac{\hbar^2}{6M}\mbox{diag}\left(\frac{k_x^2}{2},\frac{k_y^2}{3} -\frac{k^2}{6}\right).
\end{equation}
Thus it can be compensated by the respective constant external potentials for the spinor components.

Including inter-atomic interaction, averaging over the cross-section of 
the trap in the ($y,z$) plane, and rescaling variables such that the longitudinal 
coordinate is measured in the units of $a_\bot$, while the energy is measured 
in the units of  $\hbar\omega_\bot$ (where $\omega_\bot$ is the linear harmonic 
oscillator frequency of the parabolic trap in the transverse direction), 
one ends up with equation (1) from the main text. 

As the final step we take into account that the experimental \ES{length scale} values of the 
coupling field are typically hundreds of nanometers. On the other hand, 
typical transverse scale of the trap is of 
several microns, while its length can be of a few hundreds of microns. Thus, 
the suggested beam configuration can create almost arbitrary, in particular random, 
potentials $\theta(x)$ using monochromatic beams, as describes above.

\end{document}